\documentclass[prl,aps,floats,letterpaper,floatfix,nofootinbib,superscriptaddress,preprintnumbers,twocolumn]{revtex4}
\usepackage{amsmath}
\usepackage{color}
\usepackage[mathenv]{}
\usepackage{graphicx}
\usepackage{natbib}

\begin{document}
\bibliographystyle{amsplain}

\newcommand{\oldtext}[1]{\textcolor{red}{\tt #1}}
\newcommand{\newtext}[1]{\textcolor{blue}{\sl #1}}

\title{An all-optical trap for a gram-scale mirror}

\author{Thomas Corbitt}
\affiliation{LIGO Laboratory, Massachusetts Institute of Technology,
Cambridge, MA 02139, USA}
\author{Yanbei Chen}
\affiliation{Max-Planck-Institut f\"{u}r Gravitationsphysik, Am
M\"{u}hlenberg 1, 14476 Potsdam, Germany}
\author{Edith Innerhofer}
\affiliation{LIGO Laboratory, Massachusetts Institute of Technology,
Cambridge, MA 02139, USA}
\author{Helge M\"{u}ller-Ebhardt}
\affiliation{Max-Planck-Institut f\"{u}r Gravitationsphysik (Albert
Einstein Institute), Callinstra{\ss}e 38, 30167 Hannover, Germany}
\author{David Ottaway}
\affiliation{LIGO Laboratory, Massachusetts Institute of Technology,
Cambridge, MA 02139, USA}
\author{Henning Rehbein}
\affiliation{Max-Planck-Institut f\"{u}r Gravitationsphysik (Albert
Einstein Institute), Callinstra{\ss}e 38, 30167 Hannover, Germany}
\author{Daniel Sigg}
\affiliation{LIGO Hanford Observatory, Route 10, Mile marker 2,
Hanford, WA 99352, USA}
\author{Stanley Whitcomb}
\affiliation{LIGO Laboratory, California Institute of Technology,
Pasadena, CA 91125, USA}
\author{Christopher Wipf}
\affiliation{LIGO Laboratory, Massachusetts Institute of Technology,
Cambridge, MA 02139, USA}
\author{Nergis Mavalvala}
\affiliation{LIGO Laboratory, Massachusetts Institute of Technology,
Cambridge, MA 02139, USA}

\begin{abstract}
  We report on a stable optical trap suitable for a macroscopic mirror,
  wherein the dynamics of the mirror are fully dominated by radiation
  pressure. The technique employs two frequency-offset laser fields
  to simultaneously create a stiff optical restoring force and a
  viscous optical damping force. We show how these forces may be used
  to optically trap a free mass without introducing thermal noise; and
  we demonstrate the technique experimentally with a 1 gram mirror.
  The observed optical spring has an inferred Young's modulus of 1.2
  TPa, 20\% stiffer than diamond. The trap is intrinsically cold and
  reaches an effective temperature of 0.8 K, limited by technical
  noise in our apparatus.
\end{abstract}

\definecolor{purple}{rgb}{0.6,0,1}
\preprint{\large \color{purple}{LIGO-P060025-00-R}}
\date{\today}
\maketitle

The change in dynamics caused by radiation pressure effects has been
explored in many mechanical systems; its proposed applications include
cooling toward the ground state of nano- or micro-electromechanical
systems (N/MEMS)~\cite{Karrai, schwabNature2006, zeilingerNature2006,
  klecknerNature2006, arcizetNature2006, vahala}, enhancing the
sensitivity of gravitational wave (GW) detectors~\cite{BC3, 40mOS},
and generation of ponderomotively squeezed
light~\cite{corbittPRA2006}. Two types of radiation pressure effects
are evident in these systems: the optical restoring and viscous
damping forces, both of which are generated by detuned optical
cavities.  Detuning a cavity to higher frequencies (blue-detuning)
gives rise to a restoring force, known as an optical
spring~\cite{BCcontrols, msupla2002os, Sheard}, as well as an
anti-damping force due to the delay in the cavity response time.
Conversely, detuning to lower frequencies (red-detuning) gives rise
to optical damping~\cite{manciniPRL1998} along with an
anti-restoring force.

In N/MEMS, optical (anti-)restoring forces are typically negligible
in comparison to the stiff mechanical suspension.  However, optical
damping produces cooling in a red-detuned cavity, while anti-damping
heats, or even leads to instability in a blue-detuned
cavity~\cite{cohadonPRL1999, Karrai, bouwmeesterPRL2006,
schwabNature2006,zeilingerNature2006,klecknerNature2006,arcizetNature2006}.
In GW detectors, on the other hand, the optical spring force may
dominate, since the mechanical suspension of their mirrors is very
soft. The typical use of the optical spring effect in these systems
is to enhance the sensitivity of the detector around the optical
spring resonance.  To achieve a restoring force, the cavity must be
blue-detuned, and the coincident optical anti-damping force can both
destabilize the cavity and give rise to parametric instabilities of
the internal modes of its
mirrors~\cite{corbittPIOS,40mOS,kippenbergPRL2005}. In general,
whenever the radiation pressure of a single optical field dominates
both the mechanical damping and restoring forces, the system is
unstable due to the presence of a strong anti-damping or
anti-restoring optical force. Hence, until now this regime has been
achieved only with the help of active feedback control to stabilize
the dynamics~\cite{corbittPIOS,40mOS}.

Here we propose and demonstrate a technique that circumvents the
optomechanical instability by using the radiation pressure of a
second optical field, thus creating a stable optical trap for a 1
gram mirror. This opens a new route to mitigating parametric
instabilities in GW detectors, and probing for quantum effects in
macroscopic objects.

\begin{figure}[t]
\begin{center}
\includegraphics[width=8cm]{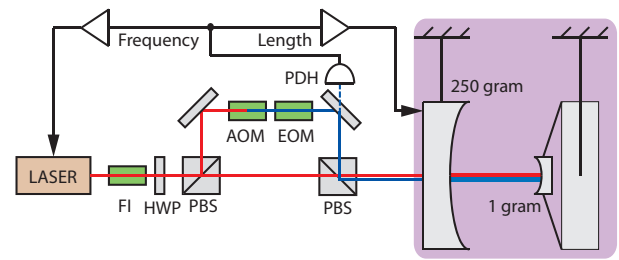}
  \caption{(Color online) Simplified schematic of the experiment. About $3$ W of
    $\lambda_0=1064$ nm Nd:YAG laser light passes through a Faraday
    isolator (FI) before it is split into two paths by a
    half-waveplate (HWP) and polarizing beamsplitter (PBS) combination
    that allows control of the laser power in each path. The carrier
    (C) field comprises most of the light incident on the suspended
    cavity. About $5$\% of the light is frequency-shifted by one free
    spectral range (161.66 MHz) using an acousto-optic modulator
    (AOM), and phase modulated by an electro-optic modulator (EOM);
    this subcarrier (SC) field can further be detuned from resonance
    to create a second optical spring. The two beams are recombined on
    a second PBS before being injected into the cavity, which is
    mounted on a seismic isolation platform in a vacuum chamber
    (denoted by the shaded box).  A Pound-Drever-Hall (PDH) error
    signal derived from the SC light reflected from the cavity is used
    to lock it, with feedback to both the cavity length as well as the
    laser frequency. By changing the frequency shift of the
    SC, the C can be shifted off resonance by
    arbitrarily large detunings. The low power SC beam (blue) passes
    through the EOM and AOM before being recombined with the high power
    C beam (red)..\label{fig:schematic}}
\end{center}
\end{figure}

The experiment shown schematically in Fig. \ref{fig:schematic} was
performed to demonstrate the optical trapping scheme. The $250$ gram
input mirror of the $L=0.9$ m long cavity is suspended as a pendulum
with oscillation frequency of $1$ Hz for the longitudinal mode. The
$1$ gram end mirror is suspended by two optical fibers $300$~$\mu {\rm
  m}$ in diameter, giving a natural frequency $\Omega_m = 2\,\pi
\times 172$~Hz for its mechanical mode, with quality factor $Q_m =
3200$.  On resonance, the intracavity power is enhanced relative to
the incoming power by a resonant gain factor $4/\mathcal{T}_i \approx
5 \times 10^3$, where $\mathcal{T}_i$ is the transmission of the input
mirror, and the resonant linewidth (HWHM) is $\gamma =
\frac{\mathcal{T}_i\,c}{4\,L} \approx 2\,\pi \times 11$ kHz.

\begin{figure}[t]
\begin{center}
\includegraphics[width=9cm]{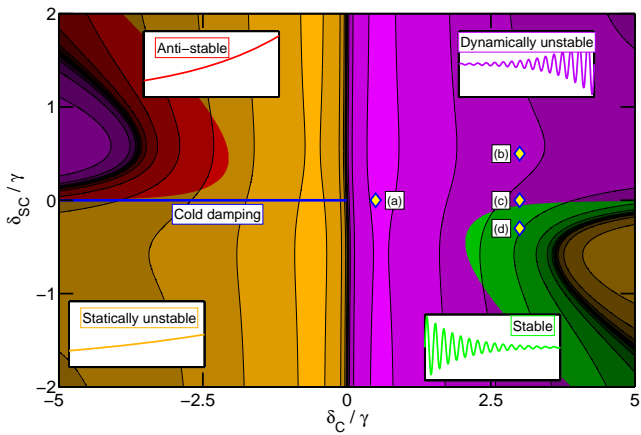}
  \caption{(Color online) Graphical representation of the total optical rigidity due
    to both optical fields, as a function of C and SC detuning, for
    fixed input power (power in the SC field is $\sim 1/20$ the C
    power) and observation frequency ($\Omega = 2\,\pi \times 1$~kHz).
    The shaded regions correspond to detunings where the total spring
    constant $K$ and damping constant $\Gamma$ are differently
    positive or negative. Specifically, ``stable'' corresponds to $K>0\,,\Gamma<0$, ``anti-stable'' to $K<0\,,\Gamma>0$, ``statically unstable'' to $K<0\,,\Gamma<0$, and
``dynamically unstable'' to $K>0\,,\Gamma>0$. The blue line denotes
``cold damping'' corresponding to $\delta_C < 0$ and $\delta_{SC} =
0$, i.e., the SC provides no optical force. The (logarithmically
spaced) contours shown
    are scaled according to $K$: brighter regions have larger $K$. The
    labels (a) -- (d) refer to the measurements shown in
    Fig.~\ref{fig:OS}.\label{fig:stability}}
\end{center}
\end{figure}

If the resonance condition is exactly satisfied, the intracavity power
depends quadratically on small changes in the length of the cavity. In
this case the radiation pressure is only a second-order effect for the
dynamics of the cavity.  The constant (dc) radiation pressure is
balanced through external forces; consequently, only fluctuations of
the radiation pressure are considered here.  If the cavity is detuned
from the resonance condition, the intracavity power, and therefore the
radiation pressure exerted on the mirrors, becomes linearly dependent
on the length of the cavity, analogous to a spring. The resulting
spring constant is given in the frequency domain
by~\cite{corbittPIOS,msupla2002os}
\begin{eqnarray}
\label{eq:K} K\left(\Omega\right) &=&K_0\,\frac{\left[1 + \left(\delta/\gamma\right)^2 -
\left(\Omega/\gamma \right)^2 \right] }{\left[1 + \left(\delta/\gamma\right)^2-\left(\Omega/\gamma\right)^2 \right]^2 +
4\,\left(\Omega/\gamma\right)^2 } \nonumber\\
\label{eq:K0} K_0 &=& \frac{2}{c}\frac{dP}{dL}=\frac{128\,\pi \,
I_0\, \left(\delta/\gamma \right) }{\mathcal{T}_i^2 \,c \, \lambda_0
}\left[\frac{1}{1+\left(\delta/\gamma\right)^2} \right],
\end{eqnarray}
where $\Omega$ is the frequency of the motion, and $\delta$ and $I_0$
are the detuning and input power of the laser, respectively.  Note the
dependence of $K_0$ on the sign of $\delta$. For $\delta>0$ (in our
convention), $K>0$ corresponds to a restoring force, while $\delta<0$
gives an anti-restoring force; we do not explore this regime
experimentally since it is always unstable for our system (see
Fig.~\ref{fig:stability}). The light in the cavity (for $\delta \ll
\gamma$) responds to mirror motion on a time scale given by
$\gamma^{-1}$. This delay has two effects. First, for high frequency
motion ($\Omega \gtrsim \gamma$), the response of the cavity, and the
corresponding radiation pressure, are reduced, and we see from
Eq.~(\ref{eq:K}) that $K\left(\Omega\gg\gamma\right) \approx
K_0\left(\Omega/\gamma\right)^{-2}$. Second, the response of the
cavity lags the motion, leading to an additional force proportional to
the velocity of the mirror motion --- a viscous force with damping
coefficient given by~\cite{corbittPIOS,msupla2002os}
\begin{eqnarray}
\label{eq:damp} \Gamma\left(\Omega\right) \equiv
\frac{2\,K\left(\Omega\right)}{M \, \gamma\left[1 + \left(\delta/\gamma\right)^2 -
\left(\Omega/\gamma \right)^2 \right]},
\end{eqnarray}
where $M$ is the reduced mass of the two mirrors. Because the cavity
response lags the motion of the mirrors, a restoring spring constant
implies a negative damping.  Again we see that when both optical
forces dominate their mechanical counterparts, the system must be
unstable.

To stabilize the system we use two optical fields that respond on
different time scales. One field should respond quickly, so that it
makes a strong restoring force and only a weak anti-damping force. The
other field should respond slowly, so that it creates a strong damping
force, with only a minor anti-restoring force. This could be achieved
with two cavities of differing bandwidths that share a common end
mirror.  However, it is simpler to use a single cavity and two fields
with vastly different detunings.  From Eqs. (\ref{eq:K}) and
(\ref{eq:damp}), taking $\Omega \ll \gamma$ (valid at the optical
spring resonant frequency), we find
\begin{equation}
  \frac{\Gamma}{K} =
  \frac{2/\left(M\,
  \gamma\right)}{1+\left(\delta/\gamma\right)^2}\,;
\end{equation}
we see that an optical field with larger detuning has less damping
per stiffness. The physical mechanism for this is that at larger
detunings, the optical field resonates less strongly than for
smaller detunings, so the time scale for the cavity response is
shorter, leading to smaller optical damping. To create a stable
system, we consider a carrier field (C) with large detuning
$\delta_C \approx 3\,\gamma$ that creates a restoring force, but
also a small anti-damping force. To counteract the anti-damping, a
strong damping force is created by injecting a subcarrier (SC) with
small detuning $\delta_{SC} \approx - 0.5 \,\gamma$. For properly
chosen power levels in each field, the resulting system is stable;
we found a factor of 20 higher power in the carrier to be suitable
in this case.  To illustrate the behavior of the system at all
detunings, the various stability regions are shown in
Fig.~\ref{fig:stability} for this fixed power ratio. Point (d) in
particular shows that the system is stable for our chosen
parameters.

\begin{figure}[t]
\begin{center}
\includegraphics[height=6cm]{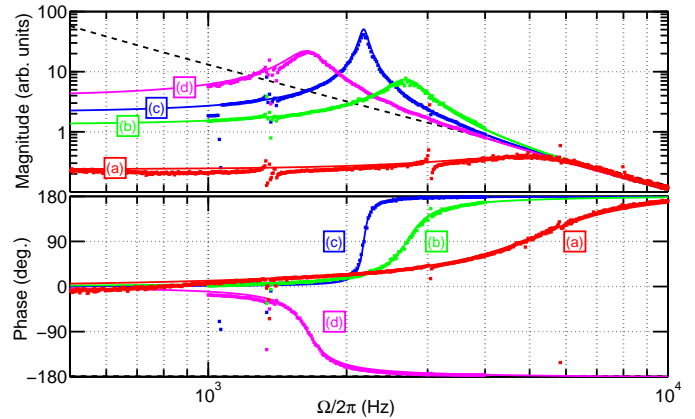}
  \caption{The optical spring response for various power levels and
    detunings of the carrier and subcarrier. Measured transfer
    functions of displacement per force are shown as points, while the
    solid lines are theoretical curves. The dashed line shows the
    response of the system with no optical spring. An unstable optical
    spring resonance with varying damping and resonant frequency is
    produced when (a) $\delta_{C}= 0.5\,\gamma\,, \delta_{SC} = 0$;
    (b) $\delta_C = 3\,\gamma\,,\delta_{SC} = 0.5\,\gamma$; (c)
    $\delta_{C} = 3\,\gamma\,,\delta_{SC} = 0$; and it is stabilized
    in (d) $\delta_{C} = 3\,\gamma\,,\delta_{SC} = -0.3\,\gamma$. Note
    that the damping of the optical spring increases greatly as the
    optomechanical resonance frequency increases, approaching
    $\Gamma_{\rm eff} \approx \Omega_{\rm eff}$ for the highest
    frequency optical spring.\label{fig:OS}}
\end{center}
\end{figure}

Next we highlight some notable features of this optical trapping
technique that were demonstrated experimentally using the apparatus
of Fig.~\ref{fig:schematic}.

(i) {\bf Extreme rigidity}: With no SC detuning and $\delta_C \approx
0.5 \,\gamma$, the $172$ Hz mechanical resonance of the $1$ gram
mirror oscillator was shifted as high as $5$ kHz (curve (a) in
Fig.~\ref{fig:OS}), corresponding to an optical rigidity of $K = 2
\times 10^6$ N/m. To put this number into perspective, consider
replacing the optical mode with a rigid beam with Young's modulus $E$.
The effective Young's modulus of this mode with area $A$ of the beam
spot ($1.5 {\rm ~mm}^2$) and length $L = 0.9$~m of the cavity, is
given by $E = K \,L / A = 1.2$ TPa, stiffer than any known material
(but also with very small breaking strength). Such rigidity is
required to operate the cavity without external control; ambient
motion would otherwise disrupt the cavity resonance condition.

(ii) {\bf Stabilization}: Also shown in Fig.~\ref{fig:OS} are curves
corresponding to various C and SC detunings. In curves (b), (c) and
(d), we detune the carrier by more than the cavity linewidth since
the optical spring is less unstable for large $\delta_C$. With no SC
detuning, the optomechanical resonant frequency reaches $\Omega_{\rm
  eff} = 2\,\pi \times 2178$ Hz, shown in curve (c). Note that the
optical spring is unstable, as evidenced by the phase {\it increase}
of $180^{\circ}$ about the resonance (corresponding to anti-damping).
Next we detune the subcarrier in the same direction as the carrier,
shown in curve (b), which increases the resonant frequency and also
increases the anti-damping, demonstrated by the broadening of the
resonant peak.  For these two cases, electronic servo control is used
to keep the cavity locked. If the control system is disabled, the
amplitude of the cavity field and mirror oscillations grow
exponentially. Remarkably, when the subcarrier is detuned in the
opposite direction from the carrier, the optical spring resonance
becomes stable, shown in curve (d), allowing operation of the cavity
without electronic feedback at frequencies above 30 Hz; we note the
change in phase behavior and the reduction of the resonant frequency.
This shows how the frequency and damping of the optical spring can be
independently controlled in the strong coupling regime.

(iii) {\bf Optical cooling}: The thermal excitation spectrum of the
mirror, given by $S_{F} = 4 k_B T \Gamma_m/M$, is not changed by the
optical forces. It is informative to express this in terms of the
optomechanical parameters $\Gamma_{\rm eff}$, $\Omega_{\rm eff}$ and
an effective temperature, $T_{\rm eff}$, such that the form of the
equation is maintained. The effective temperature thus is given by
\begin{equation}
 T_{\rm eff} = T \frac{\Gamma_m}{\Gamma_{\rm eff}} = T\, \frac{\Omega_m}{\Omega_{\rm eff}} \frac{Q_{\rm eff}}{Q_m},
\end{equation}
where $Q_i = \Omega_i/\Gamma_i$ ($i = {\rm m, eff}$) is the quality
factor of the oscillator. In the standard cold damping technique
lower $T_{\rm eff}$ is achieved by decreasing $Q_{\rm eff}$ via the
viscous radiation pressure damping. The optical spring effect
results in further cooling by increasing the resonant frequency. The
combination of both effects allows for much colder temperatures to
be attained than with cold damping alone. This is relevant to
experiments hoping to observe quantum effects in macroscopic
objects, since it greatly reduces the thermal occupation number
\begin{equation}
  N = \frac{k_{B} T_{\rm eff} } {\hbar \Omega_{\rm eff}},
\end{equation}
both by decreasing the effective temperature, and increasing the
resonant frequency.

\begin{figure}[t]
\begin{center}
\includegraphics[width=9.0cm]{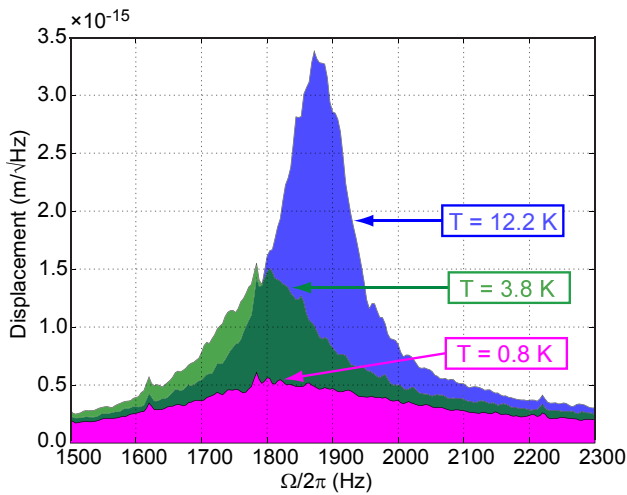}
  \caption{(Color online) The measured noise spectral density of the cavity length is
    shown for several configurations corresponding to different
    detunings. The lowest amplitude (magenta) curve corresponds to
    $\delta_C \approx 3$ and $\delta_{SC} \approx -0.5$. The other
    (green and blue) curves are obtained by reducing $\delta_{SC}$ and
    increasing$\delta_C$ in order to keep $\Omega_{\rm eff}$ approximately
    constant, while varying $\Gamma_{\rm eff}$. The spectrum is integrated between
    1500 and 2300 Hz to calculate the rms motion of the oscillator
    mode, giving effective temperatures of $0.8$, $3.8$ and $12.2$ K.
    The limiting noise source here is not thermal noise, but in fact
    frequency noise of the laser, suggesting that with reduced
    frequency noise, even lower temperatures could be attained.
    \label{fig:cool}}
\end{center}
\end{figure}

In the current experiment the displacement spectrum is dominated by
laser frequency noise at $\Omega_{\rm eff}$. We can nonetheless
estimate the effective temperature of the optomechanical mode by
measuring the displacement of the mirror, and equating ${1 \over 2} K
x_{\rm rms}^2 = {1 \over 2} k_{B} T_{\rm eff}$, where $x_{\rm rms}$ is
the RMS motion of the mirror.  To determine $x_{\rm rms}$ in our
experiment, we measure the noise spectral density of the error signal
from the cavity, calibrated by injecting a frequency modulation of
known amplitude at 12 kHz. The displacement noise measured in this way
is shown in Fig.~\ref{fig:cool}. The lowest measured temperature of
$0.8$ K corresponds to a reduction in $N$ by a factor of
$2.5\times10^3$.

In conclusion, we have exhibited a scheme that uses both the optical
spring effect and optical damping from two laser fields to create a
stable optomechanical system in which the dynamics are determined by
radiation pressure alone. We experimentally demonstrated that the
system is indeed stable, confirmed by deactivating the electronic
control system and permitting the cavity to evolve freely at the
dynamically relevant frequencies. We believe this is a useful
technique for manipulating the dynamics of radiation pressure
dominated systems, to quell their instabilities and examine their
quantum behavior free from external control.

We would like to thank our colleagues at the LIGO Laboratory and the
MQM group, and also Steve Girvin; a casual discussion with Steve
prompted some of the experimental work explored in this manuscript.
We are also grateful to Mike Boyle for first prompting us to explore
double optical springs, and to Kentaro Somiya for helpful
discussions and introduction to the MQM group. We gratefully
acknowledge support from National Science Foundation grants
PHY-0107417 and PHY-0457264. Y.C.'s research was supported by the
Alexander von Humboldt Foundation's Sofja Kovalevskaja Award (funded
by the German Federal Ministry of Education and Research).

\end{document}